\RequirePackage{lineno}
\documentclass[aps,prd,twocolumn,superscriptaddress,showpacs]{revtex4}
\usepackage{epsfig}
\usepackage[mediumspace,mediumqspace,squaren]{SIunits}
 \newcommand{\sysmte}{18}
 \newcommand{\sysmtm}{16}
 
 \newcommand{\syspte}{19}
 \newcommand{\sysptm}{18}
 
 \newcommand{\sysmete}{22}
 \newcommand{\sysmetm}{20}

\newcommand{\blueall}{80~387}
\newcommand{\bluealle}{19}
\newcommand{\blueallc}{6.6}

\newcommand{\syspscalecombin}{7}

\newcommand{\sysholecombin}{2}

\newcommand{\sysbkgcombin}{3}
\newcommand{\sysgtwocombin}{5}
\newcommand{\sysqedcombin}{4}
\newcommand{\syspdfcombin}{10}
\newcommand{\statcombin}{12}
\def\met{\mbox{${\hbox{$E$\kern-0.6em\lower-.1ex\hbox{/}}}_T$}} 
\def\mex{\mbox{${\hbox{$E$\kern-0.6em\lower-.1ex\hbox{/}}}_x$}} 
\def\mey{\mbox{${\hbox{$E$\kern-0.6em\lower-.1ex\hbox{/}}}_y$}} 
\def\mexy{\mbox{${\hbox{$E$\kern-0.6em\lower-.1ex\hbox{/}}}_{x,y}$}} 
 
\def\D0{D\O}                            
\begin{document}
\lefthyphenmin=2
\righthyphenmin=3

%
%

\title{Precise measurement of the $W$-boson mass with the CDF II detector}

\affiliation{Institute of Physics, Academia Sinica, Taipei, Taiwan 11529, Republic of China}
\affiliation{Argonne National Laboratory, Argonne, Illinois 60439, USA}
\affiliation{University of Athens, 157 71 Athens, Greece}
\affiliation{Institut de Fisica d'Altes Energies, ICREA, Universitat Autonoma de Barcelona, E-08193, Bellaterra (Barcelona), Spain}
\affiliation{Baylor University, Waco, Texas 76798, USA}
\affiliation{Istituto Nazionale di Fisica Nucleare Bologna, $^{ee}$University of Bologna, I-40127 Bologna, Italy}
\affiliation{University of California, Davis, Davis, California 95616, USA}
\affiliation{University of California, Los Angeles, Los Angeles, California 90024, USA}
\affiliation{Instituto de Fisica de Cantabria, CSIC-University of Cantabria, 39005 Santander, Spain}
\affiliation{Carnegie Mellon University, Pittsburgh, Pennsylvania 15213, USA}
\affiliation{Enrico Fermi Institute, University of Chicago, Chicago, Illinois 60637, USA}
\affiliation{Comenius University, 842 48 Bratislava, Slovakia; Institute of Experimental Physics, 040 01 Kosice, Slovakia}
\affiliation{Joint Institute for Nuclear Research, RU-141980 Dubna, Russia}
\affiliation{Duke University, Durham, North Carolina 27708, USA}
\affiliation{Fermi National Accelerator Laboratory, Batavia, Illinois 60510, USA}
\affiliation{University of Florida, Gainesville, Florida 32611, USA}
\affiliation{Laboratori Nazionali di Frascati, Istituto Nazionale di Fisica Nucleare, I-00044 Frascati, Italy}
\affiliation{University of Geneva, CH-1211 Geneva 4, Switzerland}
\affiliation{Glasgow University, Glasgow G12 8QQ, United Kingdom}
\affiliation{Harvard University, Cambridge, Massachusetts 02138, USA}
\affiliation{Division of High Energy Physics, Department of Physics, University of Helsinki and Helsinki Institute of Physics, FIN-00014, Helsinki, Finland}
\affiliation{University of Illinois, Urbana, Illinois 61801, USA}
\affiliation{The Johns Hopkins University, Baltimore, Maryland 21218, USA}
\affiliation{Institut f\"{u}r Experimentelle Kernphysik, Karlsruhe Institute of Technology, D-76131 Karlsruhe, Germany}
\affiliation{Center for High Energy Physics: Kyungpook National University, Daegu 702-701, Korea; Seoul National University, Seoul 151-742, Korea; Sungkyunkwan University, Suwon 440-746, Korea; Korea Institute of Science and Technology Information, Daejeon 305-806, Korea; Chonnam National University, Gwangju 500-757, Korea; Chonbuk National University, Jeonju 561-756, Korea}
\affiliation{Ernest Orlando Lawrence Berkeley National Laboratory, Berkeley, California 94720, USA}
\affiliation{University of Liverpool, Liverpool L69 7ZE, United Kingdom}
\affiliation{University College London, London WC1E 6BT, United Kingdom}
\affiliation{Centro de Investigaciones Energeticas Medioambientales y Tecnologicas, E-28040 Madrid, Spain}
\affiliation{Massachusetts Institute of Technology, Cambridge, Massachusetts 02139, USA}
\affiliation{Institute of Particle Physics: McGill University, Montr\'{e}al, Qu\'{e}bec, Canada H3A~2T8; Simon Fraser University, Burnaby, British Columbia, Canada V5A~1S6; University of Toronto, Toronto, Ontario, Canada M5S~1A7; and TRIUMF, Vancouver, British Columbia, Canada V6T~2A3}
\affiliation{University of Michigan, Ann Arbor, Michigan 48109, USA}
\affiliation{Michigan State University, East Lansing, Michigan 48824, USA}
\affiliation{Institution for Theoretical and Experimental Physics, ITEP, Moscow 117259, Russia}
\affiliation{University of New Mexico, Albuquerque, New Mexico 87131, USA}
\affiliation{The Ohio State University, Columbus, Ohio 43210, USA}
\affiliation{Okayama University, Okayama 700-8530, Japan}
\affiliation{Osaka City University, Osaka 588, Japan}
\affiliation{University of Oxford, Oxford OX1 3RH, United Kingdom}
\affiliation{Istituto Nazionale di Fisica Nucleare, Sezione di Padova-Trento, $^{ff}$University of Padova, I-35131 Padova, Italy}
\affiliation{University of Pennsylvania, Philadelphia, Pennsylvania 19104, USA}
\affiliation{Istituto Nazionale di Fisica Nucleare Pisa, $^{gg}$University of Pisa, $^{hh}$University of Siena and $^{ii}$Scuola Normale Superiore, I-56127 Pisa, Italy}
\affiliation{University of Pittsburgh, Pittsburgh, Pennsylvania 15260, USA}
\affiliation{Purdue University, West Lafayette, Indiana 47907, USA}
\affiliation{University of Rochester, Rochester, New York 14627, USA}
\affiliation{The Rockefeller University, New York, New York 10065, USA}
\affiliation{Istituto Nazionale di Fisica Nucleare, Sezione di Roma 1, $^{jj}$Sapienza Universit\`{a} di Roma, I-00185 Roma, Italy}
\affiliation{Rutgers University, Piscataway, New Jersey 08855, USA}
\affiliation{Texas A\&M University, College Station, Texas 77843, USA}
\affiliation{Istituto Nazionale di Fisica Nucleare Trieste/Udine, I-34100 Trieste, $^{kk}$University of Udine, I-33100 Udine, Italy}
\affiliation{University of Tsukuba, Tsukuba, Ibaraki 305, Japan}
\affiliation{Tufts University, Medford, Massachusetts 02155, USA}
\affiliation{University of Virginia, Charlottesville, Virginia 22906, USA}
\affiliation{Waseda University, Tokyo 169, Japan}
\affiliation{Wayne State University, Detroit, Michigan 48201, USA}
\affiliation{University of Wisconsin, Madison, Wisconsin 53706, USA}
\affiliation{Yale University, New Haven, Connecticut 06520, USA}

\author{T.~Aaltonen}
\affiliation{Division of High Energy Physics, Department of Physics, University of Helsinki and Helsinki Institute of Physics, FIN-00014, Helsinki, Finland}
\author{B.~\'{A}lvarez~Gonz\'{a}lez$^z$}
\affiliation{Instituto de Fisica de Cantabria, CSIC-University of Cantabria, 39005 Santander, Spain}
\author{S.~Amerio}
\affiliation{Istituto Nazionale di Fisica Nucleare, Sezione di Padova-Trento, $^{ff}$University of Padova, I-35131 Padova, Italy}
\author{D.~Amidei}
\affiliation{University of Michigan, Ann Arbor, Michigan 48109, USA}
\author{A.~Anastassov$^x$}
\affiliation{Fermi National Accelerator Laboratory, Batavia, Illinois 60510, USA}
\author{A.~Annovi}
\affiliation{Laboratori Nazionali di Frascati, Istituto Nazionale di Fisica Nucleare, I-00044 Frascati, Italy}
\author{J.~Antos}
\affiliation{Comenius University, 842 48 Bratislava, Slovakia; Institute of Experimental Physics, 040 01 Kosice, Slovakia}
\author{G.~Apollinari}
\affiliation{Fermi National Accelerator Laboratory, Batavia, Illinois 60510, USA}
\author{J.A.~Appel}
\affiliation{Fermi National Accelerator Laboratory, Batavia, Illinois 60510, USA}
\author{T.~Arisawa}
\affiliation{Waseda University, Tokyo 169, Japan}
% ...

\collaboration{CDF Collaboration\footnote{With visitors from
$^a$Istituto Nazionale di Fisica Nucleare, Sezione di Cagliari, 09042 Monserrato (Cagliari), Italy,
$^b$University of CA Irvine, Irvine, CA 92697, USA,
$^c$University of CA Santa Barbara, Santa Barbara, CA 93106, USA,
$^d$University of CA Santa Cruz, Santa Cruz, CA 95064, USA,
$^e$Institute of Physics, Academy of Sciences of the Czech Republic, Czech Republic,
$^f$CERN, CH-1211 Geneva, Switzerland,
$^g$Cornell University, Ithaca, NY 14853, USA,
$^h$University of Cyprus, Nicosia CY-1678, Cyprus,
$^i$Office of Science, U.S. Department of Energy, Washington, DC 20585, USA,
$^j$University College Dublin, Dublin 4, Ireland,
$^k$ETH, 8092 Zurich, Switzerland,
$^l$University of Fukui, Fukui City, Fukui Prefecture, Japan 910-0017,
$^m$Universidad Iberoamericana, Mexico D.F., Mexico,
$^n$University of Iowa, Iowa City, IA 52242, USA,
$^o$Kinki University, Higashi-Osaka City, Japan 577-8502,
$^p$Kansas State University, Manhattan, KS 66506, USA,
$^q$Korea University, Seoul, 136-713, Korea,
$^r$University of Manchester, Manchester M13 9PL, United Kingdom,
$^s$Queen Mary, University of London, London, E1 4NS, United Kingdom,
$^t$University of Melbourne, Victoria 3010, Australia,
$^u$Muons, Inc., Batavia, IL 60510, USA,
$^v$Nagasaki Institute of Applied Science, Nagasaki, Japan,
$^w$National Research Nuclear University, Moscow, Russia,
$^x$Northwestern University, Evanston, IL 60208, USA,
$^y$University of Notre Dame, Notre Dame, IN 46556, USA,
$^z$Universidad de Oviedo, E-33007 Oviedo, Spain,
$^{aa}$CNRS-IN2P3, Paris, F-75205 France,
$^{bb}$Texas Tech University, Lubbock, TX 79609, USA,
$^{cc}$Universidad Tecnica Federico Santa Maria, 110v Valparaiso, Chile,
$^{dd}$Yarmouk University, Irbid 211-63, Jordan,
}}
\noaffiliation

%
%
\begin{abstract}
We have measured the $W$-boson mass $M_W$ using data corresponding to 2.2~fb$^{-1}$ of integrated luminosity
  collected in $p\bar{p}$ collisions at $\sqrt{s}=1.96$~TeV with the CDF II detector at the Fermilab Tevatron
 collider. Samples consisting of 470~126 $W\to e\nu$ candidates and 624~708 $W\to\mu\nu$ candidates  yield the
  measurement $M_W = 80~387 \pm 12_{\rm stat} \pm 15_{\rm syst} = 80~387 \pm 19$~MeV$/c^2$. 
 This is the most precise measurement of the $W$-boson mass to date and 
 significantly exceeds the precision of all previous measurements combined. 
\end{abstract}


\maketitle

\vfill\eject

The mass of the $W$ boson, $M_W$, is an important parameter of the standard model (SM) of particle physics.  
 Precise measurements of $M_W$ and of other electroweak observables significantly constrain the mass of the as-yet unobserved Higgs boson, 
 which is predicted by the electroweak symmetry-breaking mechanism in the SM.  Previous 
 measurements~\cite{cdfd0,LEPexp,prl200pb,prd200pb}  yield a world average value of $M_W = 80~399 \pm 23$ MeV~\cite{tevewwg,coords} and, 
 in conjunction with other electroweak data, determine the Higgs boson 
 mass to be $M_H= 89^{+35}_{-26}$ GeV~\cite{lepwmass}. If the Higgs boson is observed,  the comparison of its directly-measured mass
  with the  SM prediction will be a powerful test of the model.  An exclusion of the Higgs boson in the predicted mass
 range by direct searches would decisively point to new physics beyond the SM, for example  
 radiative corrections from supersymmetric particles to $M_W$\cite{heinemeyer}. 
 
The production of $W$ bosons at $\sqrt{s} = 1.96$~TeV at the Fermilab Tevatron $p \bar{p}$ collider is dominated by
  the annihilation process $q^\prime\bar{q} \to W+X$ where $X$ is  initial-state QCD radiation. Leptonic decays of the $W$ boson,
  $W\to \ell\nu_\ell$ ($\ell=e,\mu$), provide high-purity samples that allow  a precise measurement of $M_W$. 

In this Letter we report a measurement of $M_W$ using fits to three kinematic
     distributions in $W \to \mu \nu$ and $W \to e \nu$ decays. 
 This measurement uses 
 data corresponding to an integrated luminosity of 2.2~fb$^{-1}$ of $p\bar{p}$ collisions collected by the CDF II detector between 2002 and 2007,
      and supersedes an earlier result obtained in a subset of these data~\cite{prl200pb,prd200pb}.  The CDF II detector~\cite{prd200pb} is a
  general-purpose apparatus designed to study $p\bar{p}$ collisions at the Tevatron.  In this analysis, charged-particle trajectories (tracks) 
 are reconstructed and measured using a drift chamber (COT)~\cite{cotNim} immersed in a 1.4~T solenoidal 
 magnetic field. Electromagnetic (EM) and hadronic calorimeters provide shower energy measurements as well as position measurements via 
 wire chambers embedded at the EM shower maximum. Surrounding the calorimeters, drift chambers~\cite{muondetectors} identify muon candidates.
  Events are selected online if they have a 
 muon (electron) with $p_T >18$~GeV ($E_T > 18$~GeV)~\cite{coords}.

Offline we select muon candidates defined by a COT track having $p_T>30$~GeV and associated with a minimum-ionizing energy deposition in the
  calorimeter and matching hits in the muon chambers. Cosmic rays are rejected with high efficiency  
 using COT hit timing~\cite{cotcosmic}. Electron candidates are required to have a COT track with $p_T>18$~GeV and an EM calorimeter cluster 
 with $E_T>30$~GeV and must pass quality requirements on the COT track and the track-cluster
 matching. Additionally, they must satisfy requirements on the following quantities: pseudorapidity ($|\eta| < 1$)~\cite{coords},
  the ratio of cluster energy
  to track momentum ($E/p < 1.6$), the ratio 
 of energies detected in the hadronic and EM calorimeters ($E_{\rm Had}/E_{\rm EM} < 0.1$), and a $\chi^2$-based difference between the expected
  and observed transverse shower profiles~\cite{prd200pb,ces}. We impose calorimeter fiduciality requirements on electron candidates to ensure uniformity
 of response. When selecting the $W$-boson candidate sample, we suppress the $Z$-boson background by rejecting
  events with a second lepton. Events composing control samples of $Z$-boson candidates are required to have two
 oppositely-charged leptons satisfying the above criteria and an invariant mass ($m_{\ell\ell}$) between 66 and 116 GeV
  and vector-summed $p_T$ ($p_T^{\ell\ell}$) less than 30 GeV. 
 
 We define the hadronic recoil $\vec{u} = \Sigma_i E_i \sin(\theta_i)\hat{n}_i$, where the sum is performed over
  calorimeter towers~\cite{run1detector}, with energy $E_i$, polar angle $\theta_i$, and transverse directions specified 
 by unit vectors $\hat{n}_i$. The sum excludes towers that  contain energy deposition from the charged lepton(s). From $\vec{p}_T$
  conservation, the transverse momentum of the neutrino is inferred 
 as $\vec{p}_T^{~\nu} \equiv  - \vec{p}_T^{~\ell} - \vec{u} $, where $\vec{p}_T^{~\ell}$ is
 the vector $p_T$ ($E_T$) of the muon (electron). We calculate the $W$-boson transverse mass as 
 \begin{equation}
 m_T = \sqrt {2 ~ (~p_T^\ell ~ p_T^\nu  - \vec{p}_T^{~\ell} \cdot \vec{p}_T^{~\nu}) }.
 \end{equation} 
 \noindent To obtain high-purity samples of $W$ bosons,
  we require $30 < p_T^\ell < 55$~GeV,  $30 < p_T^\nu  < 55$~GeV, 
 $|\vec{u}| < 15$~GeV, and $60<m_T<100$~GeV. The final  samples 
  consist of 470~126 (16~134) $W\to e\nu$ ($Z \to ee$) candidates and 624~708 (59~738) $W\to\mu\nu$ ($Z \to \mu \mu$) candidates.
 
 Measurements of $M_W$ are extracted by performing binned maximum likelihood fits to the observed distributions
  of $m_T$, $p_T^\ell$, and $p_T^\nu$ using
 simulated line-shapes (``templates'') as a function of $M_W$. A custom Monte Carlo simulation is
 used to generate templates between 80 GeV and 81 GeV. The simulation includes a boson production and decay model, and 
 a detailed model of detector response. The kinematics of $W$ and $Z$ boson production and decay are modeled using
  the {\sc resbos}~\cite{resbos} generator.   Using the $Z$-boson data, we tune the non-perturbative form factor in {\sc resbos}, which 
 describes the boson $p_T$ spectrum at low $p_T$ ($\sim$5~GeV), and $\alpha_s$, which describes the boson $p_T$ spectrum at 
 intermediate $p_T$ ($\sim$15~GeV). The radiation
  of multiple final-state photons is modeled with {\sc photos}~\cite{photos}. The {\sc photos} QED model was checked
  with {\sc horace}~\cite{horace}, which in addition to a leading-logarithm calculation of multiple initial- and final-state photons, also performs
  an exact $\cal{O}$($\alpha$) calculation. We use the CTEQ6.6~\cite{cteq66}  parton distribution functions 
 (PDFs) of the (anti)proton and verify that the MSTW2008~\cite{mstw2008} PDFs give consistent results. The 
 CTEQ6.6 and MSTW2008 PDFs yield similar estimates of the $M_W$ uncertainty. 
 We quote the 68\% confidence level (C.L.) uncertainty from the MSTW2008 ensemble of PDFs as a systematic uncertainty on $M_W$. 

The charged-lepton track is simulated using a detailed model of the passive material in the tracking volume and of individual 
 position measurements in the COT. We use a highly granular lookup table to model ionization and radiative energy loss, multiple 
 Coulomb scattering, and Compton scattering in the tracking volume. The simulation generates and propagates bremsstrahlung photons and 
 conversion electrons to the calorimeter and includes Landau-Pomeranchuk-Migdal~\cite{lpm} suppression for soft photon emission.
  Muon tracks from $\Upsilon, W$,  and $Z$-boson decays are used to determine the COT position measurement 
 resolution ($\approx$\unit{150}{\micro\meter}), which is implemented in the simulation as a function of radius. A helix fit 
 (with beam constraint for promptly-produced tracks) 
 is performed to simulate the reconstructed track. 

A high-purity sample of cosmic ray muons collected concurrently with the collider data is used
  to perform a precise alignment of the COT. The trajectory of each cosmic ray muon is fitted to a single helix through the entire COT. 
 This fit provides a robust reference for the internal alignment of sense wires, including gravitational and electrostatic displacements,
  resulting in a \unit{2\mbox{-}5}{\micro\meter} precision in relative wire positions. We remove the remaining weakly-constrained modes of
  COT deformation, based on the observed
 difference of $\left\langle E/p \right\rangle$ between positrons and electrons from $W$-boson decays. 

\begin{figure}[htb]
\begin{center}
\includegraphics[width=0.475\textwidth] {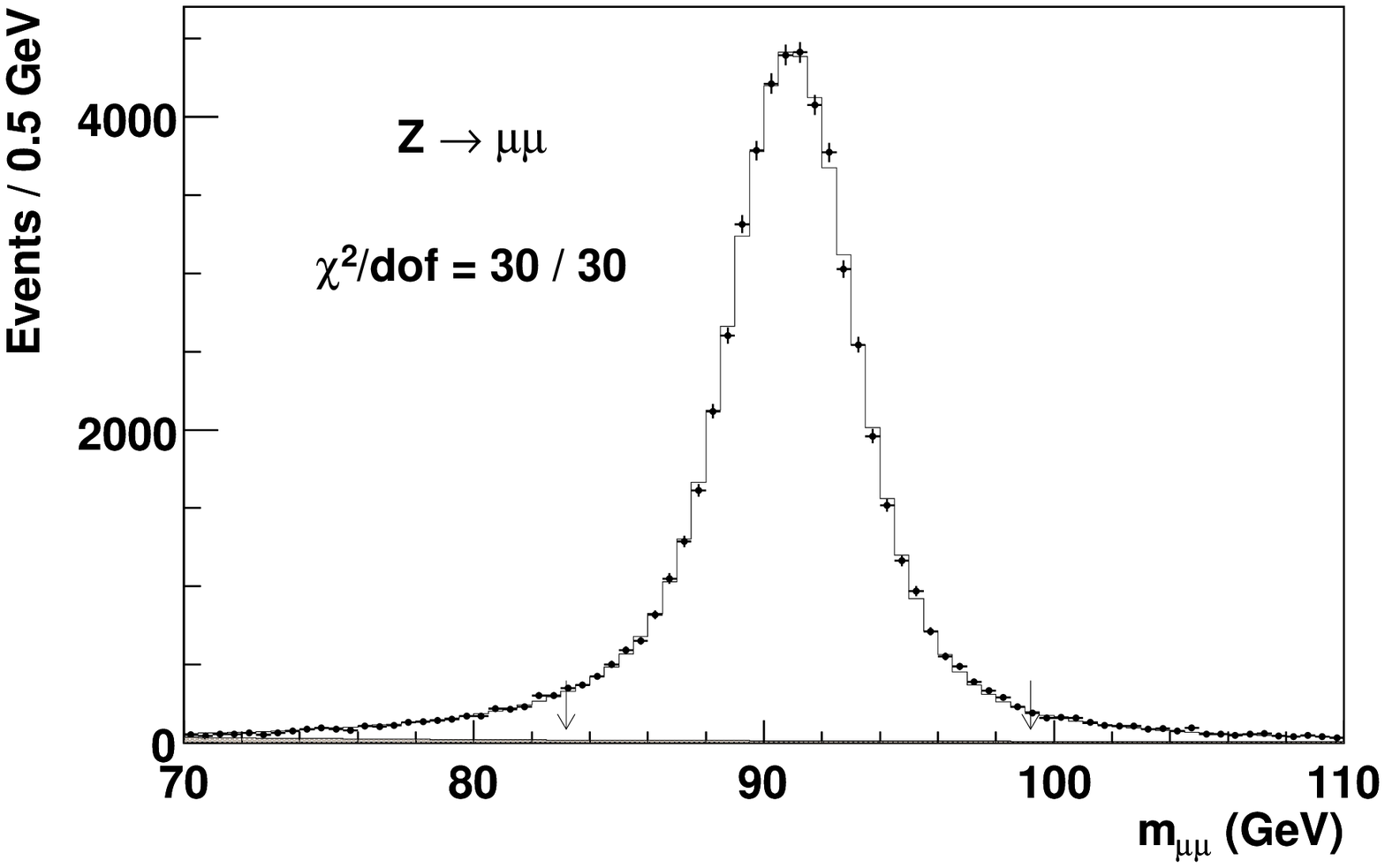}
\includegraphics[width=0.475\textwidth]{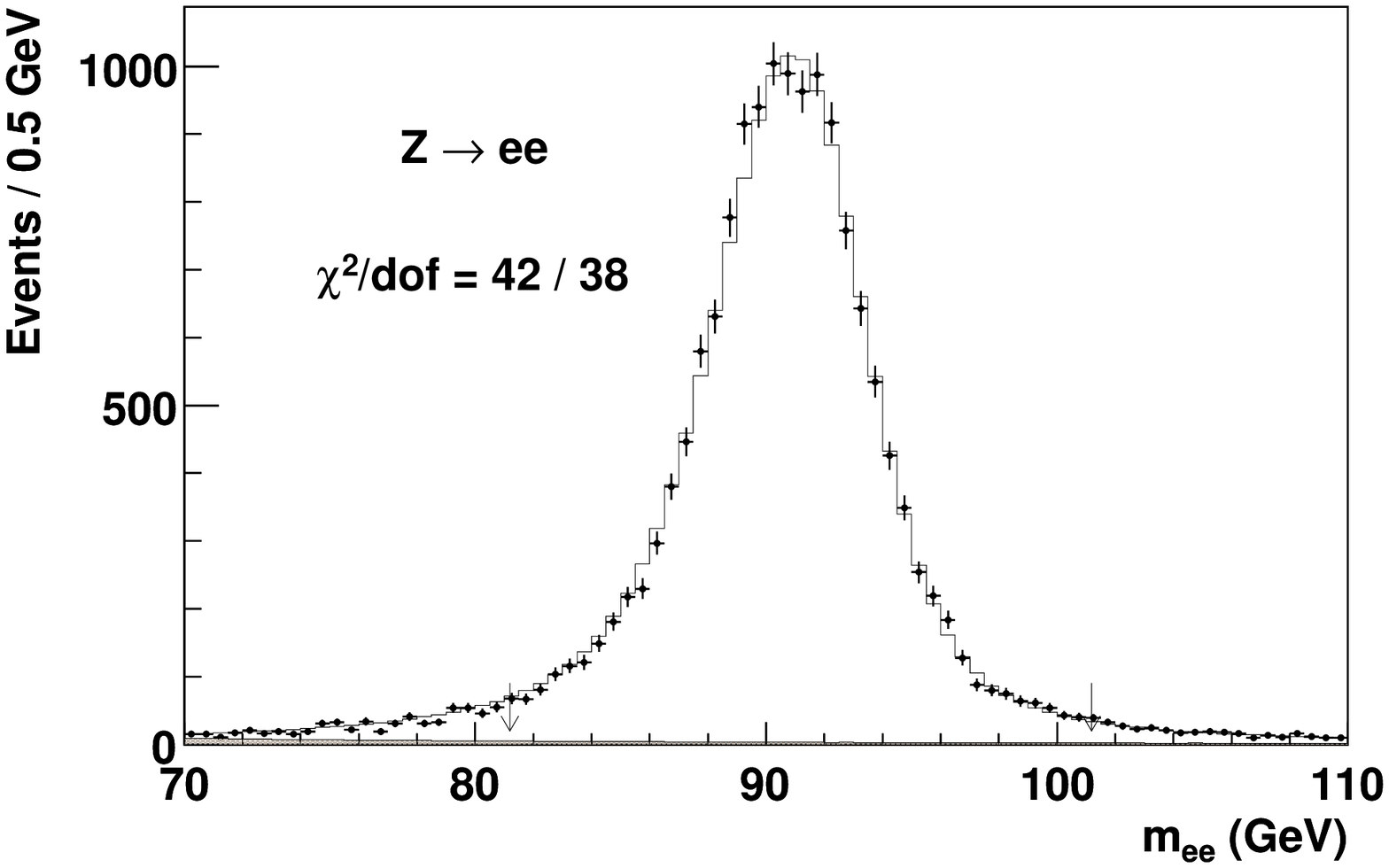}
\caption{The $Z \rightarrow \mu \mu $ (top) and $Z \rightarrow ee $ (bottom)  mass fits, showing the data (points), the best-fit  simulation template
  (histogram) and the 
 photon-pole contribution (shaded). The arrows indicate the fitting range. }
\label{z_mass_fit}
\end{center}
\end{figure}

We calibrate the tracker momentum scale using $J/\psi \rightarrow \mu \mu$ and $\Upsilon (1S) \rightarrow \mu \mu$ samples,  
 by performing a maximum-likelihood fit of the data to simulated invariant mass templates generated using the known 
 mass values of these mesons~\cite{langacker}. The momentum scale is calibrated after alignment and energy loss corrections are 
 derived from the $J/\psi$ sample. 
Nonuniformities in the tracker magnetic field are corrected by measuring the dependence of the $J/\psi$ mass on the
  mean polar angle of the muons. The dependence of the momentum scale on the difference of the muon polar angles is used to calibrate
  the polar angle measurement and the residual bias in track curvature as a function of polar angle. A  4\% correction to the
  ionization energy loss is applied to eliminate the dependence of the momentum scale on  $\langle 1/p_T \rangle$
of the muons. 

After finalizing this calibration, we perform a measurement of the $Z$-boson mass in the dimuon channel (see Fig.~\ref{z_mass_fit}), 
 initially blinded with an 
 additive offset randomly selected from a flat distribution in the range [-75,75] MeV. The unblinded result is
  $M_Z=91~180 \pm 12_{\rm stat} \pm 10_{\rm syst}$~MeV. This measurement is consistent with the world average 
 of $91~188 \pm 2$~MeV~\cite{langacker,lepwmass}, providing an incisive cross-check of the tracking simulation and the momentum scale. 
 Subsequently, we include the $Z\to\mu\mu$ 
 mass measurement as a constraint on the momentum scale.  The systematic uncertainties  due to QED radiative corrections and
  magnetic field nonuniformity dominate the total  uncertainty of 0.009\% in the combined momentum scale.

\begin{figure}[htb]
\begin{center}
\epsfysize = 4.4cm
\epsffile {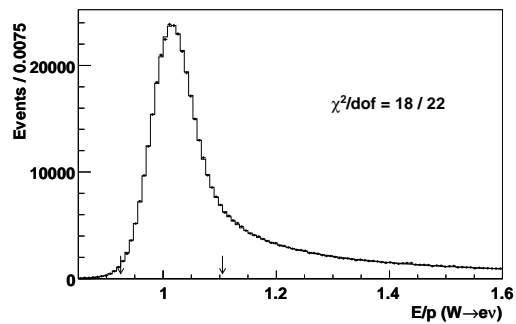}
\caption{The distribution of $E / p$ for the $W \rightarrow e \nu$ data (points) and the 
  best-fit simulation (histogram) including the small jet background. The arrows indicate the fitting range used for the
 electron energy calibration.}
\label{eop}
\end{center}
\end{figure}

In the simulation of the electron cluster, nearby bremsstrahlung photons and
  conversion electrons have their energies merged with that of the primary electron. We use
  a custom implementation of {\sc geant4}~\cite{geant} to model the distributions of electron and photon energy loss in the  solenoid coil and energy leakage into the hadronic calorimeter, as a function of $E_T$ and incident angle. Using the calibrated tracker momentum scale, we fit the $E/p$ peak in $W\to e\nu$~(Fig.~\ref{eop}) and $Z\to ee$ data in bins of $E_T$ to determine the electron energy scale and  non-linearity of the calorimeter response. We fit the radiative tail of the $E/p$ distribution to tune the amount of simulated material upstream of the COT by 2.6\%. 
 The EM calorimeter resolution is parametrized as the quadrature sum of a sampling term (12.6\%/$\sqrt{E_T/ \rm{GeV}}$) and a constant term $\kappa = (0.68 \pm 0.05)$\% applied to the cluster energy. A secondary constant term $\kappa_\gamma = (7.4\pm 1.8)$\% is applied only to the energies of bremsstrahlung photons and conversion electrons. We tune $\kappa$ on the width of the $E/p$ peak in the $W\to e\nu$ sample  and $\kappa_\gamma$ on the width of the mass peak in a $Z\to ee$ subsample where both electrons have $E/p > 1.11$. 

We use the tuned energy scale to perform an independent measurement of the $Z$-boson 
 mass in the dielectron channel (see Fig.~\ref{z_mass_fit}), initially blinded with the same offset as used for the measurement in
 the dimuon channel. The unblinded result,   
 $M_Z = 91~230\pm 30_{\rm stat} \pm 14_{\rm syst}$~MeV, is consistent with the world average, providing a stringent cross-check
 of our EM calorimeter energy scale calibration and electron simulation. Cross-checks of the $Z\to ee$ mass measurements using exclusive
  subsamples consisting of electrons with $E/p > 1.11$ and $E/p < 1.11$ respectively, 
 performed with both calorimetry and tracking,  give consistent results. The final
 determination of the electron energy scale combines the $E/p$-based calibration with the $M_Z$ measurement, taking the correlated uncertainty
  due to the QED radiative correction into account.

 The calorimeter towers containing lepton energy depositions are excluded from the calculation of the
  recoil vector $\vec{u}$. The underlying event energy in these towers is measured using the nearby towers in $W$-boson  data.
   The   $\vec{u}$ 
 resolution due to the underlying event (additional $p \bar{p}$ collisions) is modeled using data triggered on  inelastic $p\bar{p}$ interactions (random bunch crossings).  
 The  $\vec{p}_T$-imbalance between the $\vec{p}_T^{\ell\ell}$ and $\vec{u}$ 
 in $Z \rightarrow \ell \ell$ events is used to tune the recoil
  model, which also includes the response to the initial state QCD radiation and its resolution.  
 Cross-checks of the recoil model show good agreement between $W$-boson data and simulation. 

\begin{table}[htpb]
\begin{center}
\begin{ruledtabular}
\begin{tabular}{ccc}
 Distribution~  & ~ $W$-boson mass (MeV)~  &~ $\chi^2$/dof \\
\hline
 $m_T(e,\nu)$  &  $80~408 \pm 19_{\rm stat} \pm  \sysmte_{\rm syst}$   & 52/48   \\
 $p_T^\ell(e)$        &  $80~393 \pm 21_{\rm stat} \pm  \syspte_{\rm syst}$   & 60/62   \\
 $p_T^\nu(e)$       &  $80~431 \pm 25_{\rm stat} \pm  \sysmete_{\rm syst}$   & 71/62   \\
\hline
 $m_T (\mu,\nu)$ &  $80~379 \pm 16_{\rm stat} \pm  \sysmtm_{\rm syst}$   & 58/48   \\
 $p_T^\ell(\mu)$       &  $80~348 \pm 18_{\rm stat} \pm  \sysptm_{\rm syst}$   & 54/62   \\
 $p_T^\nu (\mu)$      &  $80~406 \pm 22_{\rm stat} \pm  \sysmetm_{\rm syst}$   & 79/62   \\
\end{tabular}
\end{ruledtabular}
\caption{Fit results and uncertainties
 for $M_W$. 
 The fit windows are $65-90$~GeV for the $m_T$ fit
and $32-48$~GeV for the $p_T^\ell$ and $p_T^\nu$ fits.
 The $\chi^2$ of the fit is computed using the expected 
 statistical errors on the data points. }
\label{tab:fitResults}
\end{center}
\end{table}

Kinematic distributions of background events passing the event selection cuts are included in the template fits with their estimated
 normalizations. Backgrounds arise from jets misidentified as leptons, $Z\to\ell\ell$ decays with only one reconstrcted lepton,
  $W \to \tau \nu \to \ell \nu \bar{\nu} \nu$, pion and kaon decays in flight (DIF), and cosmic rays. We estimate jet, DIF,
  and cosmic ray backgrounds from the data and $Z \to \ell \ell$ and $W\to\tau\nu$ backgrounds 
  from simulation. Background fractions for the muon (electron) datasets are evaluated to be 7.35\% (0.14\%) 
 from $Z \to \ell \ell$ decays, 0.88\% (0.93\%) from $W\to\tau\nu$ decays, 0.04\% (0.39\%) from  jets, 0.24\% from DIF, and 0.02\% from cosmic rays. 

\begin{figure}[htb]
\begin{center}
\epsfysize = 4.5cm
\epsffile {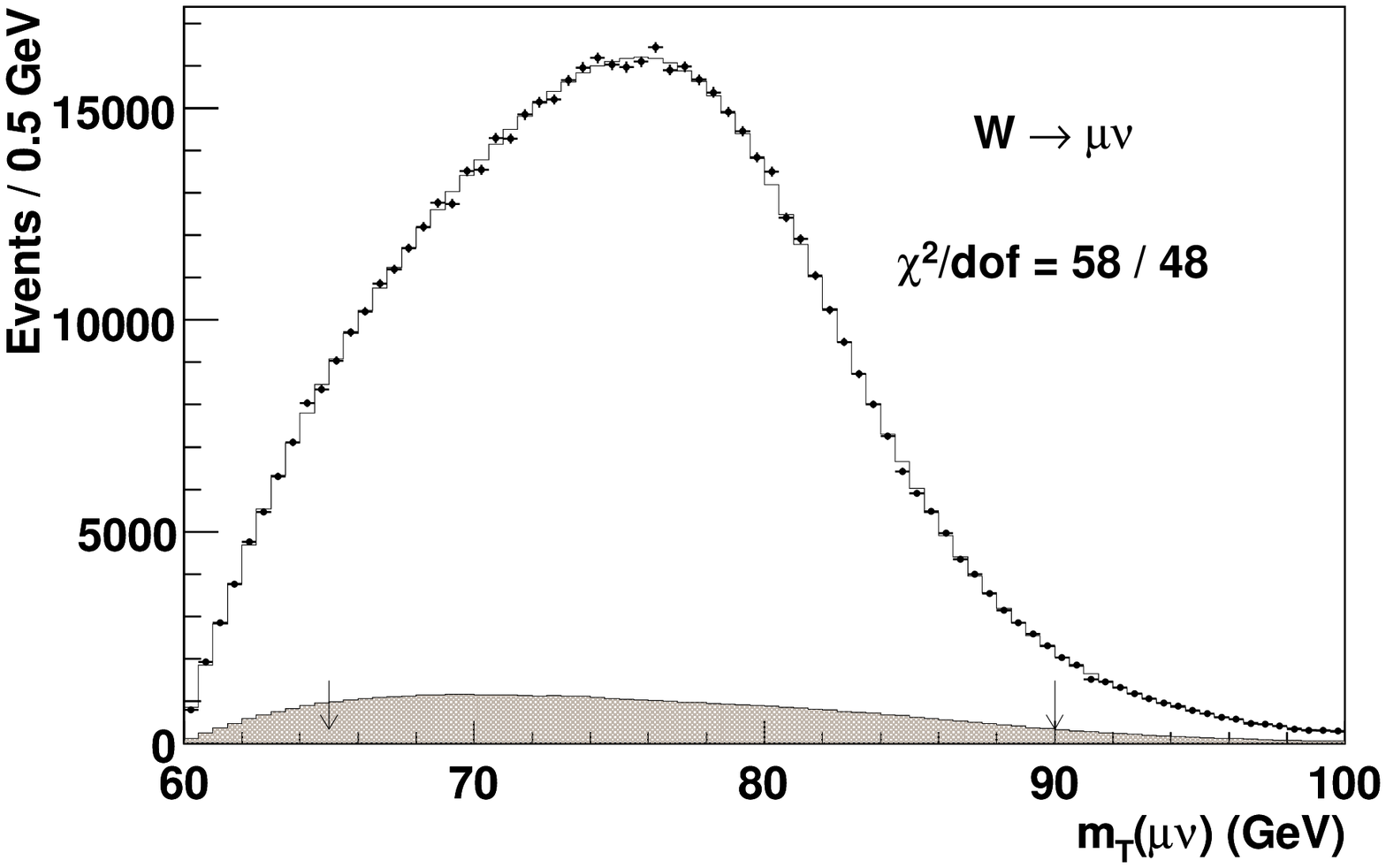}
\epsfysize = 4.5cm
\epsffile {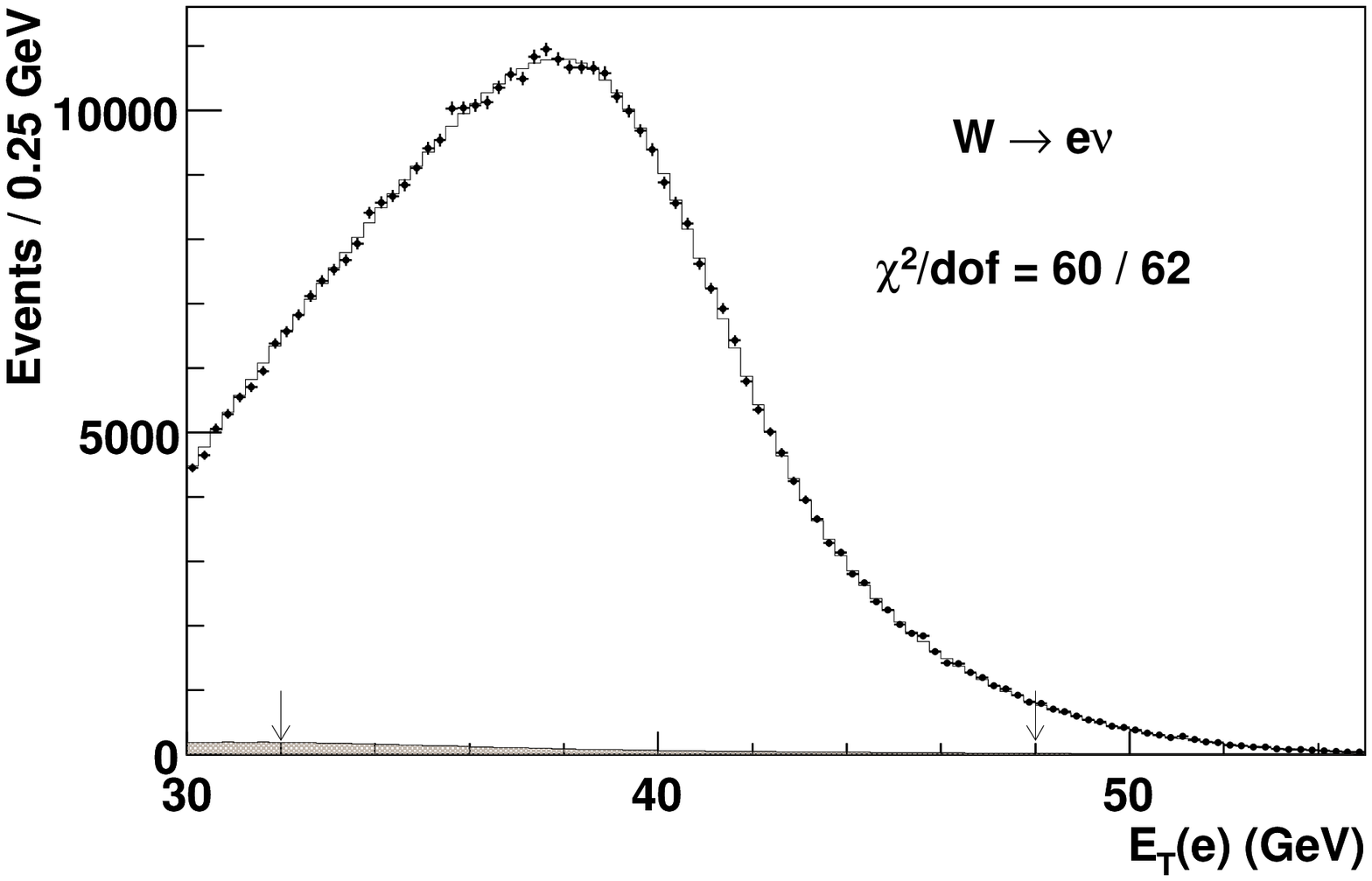}
\caption{ The $m_T$ distribution  for muons (top) and the $p_T^\ell$ distribution for electrons (bottom).  
 The data (points) and the best-fit simulation template (histogram) including backgrounds (shaded) are shown. The arrows indicate the fitting range.
}
\label{w_mass_fit}
\end{center}
\end{figure}

\begin{table}
\begin{ruledtabular}
\begin{tabular}{lr}
Source & Uncertainty (MeV)   \\ [0.05in]
\hline
Lepton energy scale and resolution             & \syspscalecombin\  \\
Recoil energy scale and resolution              &   6 \\
Lepton removal            &    \sysholecombin\    \\
Backgrounds               &    \sysbkgcombin\      \\
$p_T(W)$ model  &  \sysgtwocombin\  \\
Parton distributions      &   \syspdfcombin\   \\
QED radiation             &  \sysqedcombin\    \\
\hline
$W$-boson statistics     &    \statcombin\     \\
\hline
Total                     &    \bluealle\     \\
\end{tabular}
\caption{Uncertainties for the final combined result on $M_W$.}
\label{tab:systematics}
\end{ruledtabular}
\end{table}

The fit results (e.g., Fig.~\ref{w_mass_fit})  are summarized in Table~\ref{tab:fitResults}. As with the $Z$-boson mass measurements, the $M_W$ 
 fit values were blinded during analysis by adding another unknown offset in the range [-75,75] MeV. The consistency of these 
 results confirms  that the $W$-boson production, decay, and the hadronic recoil are well-modeled. 
 Systematic uncertainties from analysis parameters are propagated to $M_W$  by fitting events, generated with the parameter values varied
            by their uncertainties, with the nominal templates. 
 The statistical 
 correlations between fits are evaluated with simulated experiments and are found to be 69\% (68\%) between $m_T$ and $p_T^\ell$ ($p_T^\nu$) fit 
 values,  and 28\% 
 between $p_T^\ell$ and $p_T^\nu$ fit values. We perform a numerical combination of the six individually fitted $M_W$ values, including correlations, using the BLUE~\cite{bluemethod} method and obtain $M_W = \blueall \pm \bluealle$~MeV, with $\chi^2 / {\rm dof} = \blueallc / 5$. The $m_T$, $p_T^\ell$ and $p_T^\nu$ fits in the electron (muon) channel 
 contribute  weights of  17.5\% (35.5\%), 13.8\% (17.3\%), and 7.1\% (8.8\%), respectively. The systematic uncertainties  for the combined result are shown in Table~\ref{tab:systematics}. 

In conclusion, we report a new measurement of the $W$-boson mass with the CDF II detector at the Fermilab Tevatron
  using data corresponding to 2.2~fb$^{-1}$ of integrated luminosity. The measured value
 $M_W = 80~387 \pm 12_{\rm stat} \pm 15_{\rm syst} = \blueall \pm \bluealle$~MeV  is more precise than 
 all previous measurements of $M_W$ combined. The world average~\cite{tevewwg} becomes $M_W = 80~390\pm 16$ MeV.  This
  result has a significant impact on the global electroweak fit~\cite{lepwmass}; the limit on the fitted 
 mass of the SM Higgs boson has been reduced from $M_H<158$ GeV to $M_H < 145$ GeV at the 95\% C.L.

We thank the Fermilab staff and the technical staff of the participating institutions for their vital contributions. We thank 
  C. Balazs, U. Baur, C. M. Carloni Calame, K. Ellis, G. Montagna,  R. Thorne,    A. Vicini,
D. Wackeroth and Z. Was for helpful discussions. 
 This work was supported by the U.S. Department of Energy and National Science Foundation; the Italian Istituto Nazionale di Fisica Nucleare; the Ministry of Education, Culture, Sports, Science and Technology of Japan; the Natural Sciences and Engineering Research Council of Canada; the National Science Council of the Republic of China; the Swiss National Science Foundation; the A.P. Sloan Foundation; the Bundesministerium f\"ur Bildung und Forschung, Germany; the Korean World Class University Program, the National Research Foundation of Korea; the Science and Technology Facilities Council and the Royal Society, UK; the Russian Foundation for Basic Research; the Ministerio de Ciencia e Innovaci\'{o}n, and Programa Consolider-Ingenio 2010, Spain; the Slovak R\&D Agency; the Academy of Finland; and the Australian Research Council (ARC).


\begin{references}
\bibitem{coords} Pseudorapidity is defined as 
 $\eta = -\ln[\tan(\theta/2)$], where $\theta$ is the polar angle from the beam axis.  Energy (momentum)  transverse to  the beam is denoted
as $E_T$ ($p_T$). We use the convention $c = 1$ throughout this paper.
\end{references}
\end{document}